\pdfoutput=1

\documentclass[11pt]{article}
\usepackage{authblk}
\usepackage[]{naacl2021}

\usepackage[T1]{fontenc}
\usepackage[utf8]{inputenc}

\usepackage{times}
\usepackage{url}
\usepackage{latexsym}
\usepackage{enumitem}
\usepackage{color, colortbl}
\usepackage{threeparttable,booktabs}

\usepackage[hang,flushmargin]{footmisc}

\usepackage{microtype}

\usepackage{booktabs}
\usepackage{multirow}
\usepackage{adjustbox}
\usepackage{arydshln}
\usepackage{tabularx}
\usepackage{array}
\usepackage{graphicx,subcaption}
\usepackage{enumitem}
\usepackage{amsmath}
\usepackage{amsfonts}
\usepackage{amssymb}
\usepackage{dsfont}

\definecolor{bluencs}{rgb}{0.0, 0.53, 0.74}

\definecolor{Gray}{gray}{0.9}

\let\oldFootnote\footnote
\newcommand\nextToken\relax

\renewcommand\footnote[1]{%
    \oldFootnote{#1}\futurelet\nextToken\isFootnote}

\newcommand\isFootnote{%
    \ifx\footnote\nextToken\textsuperscript{,}\fi}

\title{Densifying Sparse Representations for Passage Retrieval by Representational Slicing}

\author{Sheng-Chieh Lin}
\author{Jimmy Lin}
\affil{David R. Cheriton School of Computer Science\\University of Waterloo}

\begin{document}
\maketitle
\begin{abstract}
Learned sparse and dense representations capture different successful approaches to text retrieval  and the fusion of their results has proven to be more effective and robust.
Prior work combines dense and sparse retrievers by fusing their model scores. 
As an alternative, this paper presents a simple approach to densifying sparse representations for text retrieval that does not involve any training. 
Our densified sparse representations (DSRs) are interpretable and can be easily combined with dense representations for end-to-end retrieval.
We demonstrate that our approach can jointly learn sparse and dense representations within a single model and then combine them for dense retrieval.
Experimental results suggest that combining our DSRs and dense representations yields a balanced tradeoff between effectiveness and efficiency.
\end{abstract}

\section{Introduction}
\label{sec:intro}

Transformer-based bi-encoders have been widely used as a first-stage retriever for text retrieval.
Previous work~\cite{sentence-bert, Chang2020Pre-training, dpr} trains models to project texts into a continuous embedding space for dense retrieval.
Such dense retrievers are capable of tackling the vocabulary and semantic mismatch problems compared to traditional word matching approaches (e.g., BM25).
Subsequent work further improves dense retrieval through knowledge distillation~\cite{tct, tasb}, hard negative mining~\cite{xiong2020approximate, star}, or their combination~\cite{rocketqa, condenser}.
However, as \citet{sciavolino2021simple} have shown, dense retrievers still fail in some easy cases and it remains challenging to interpret why they sometimes perform poorly.

\begin{figure}[t]
    \centering
    \resizebox{\columnwidth}{!}{
        \includegraphics{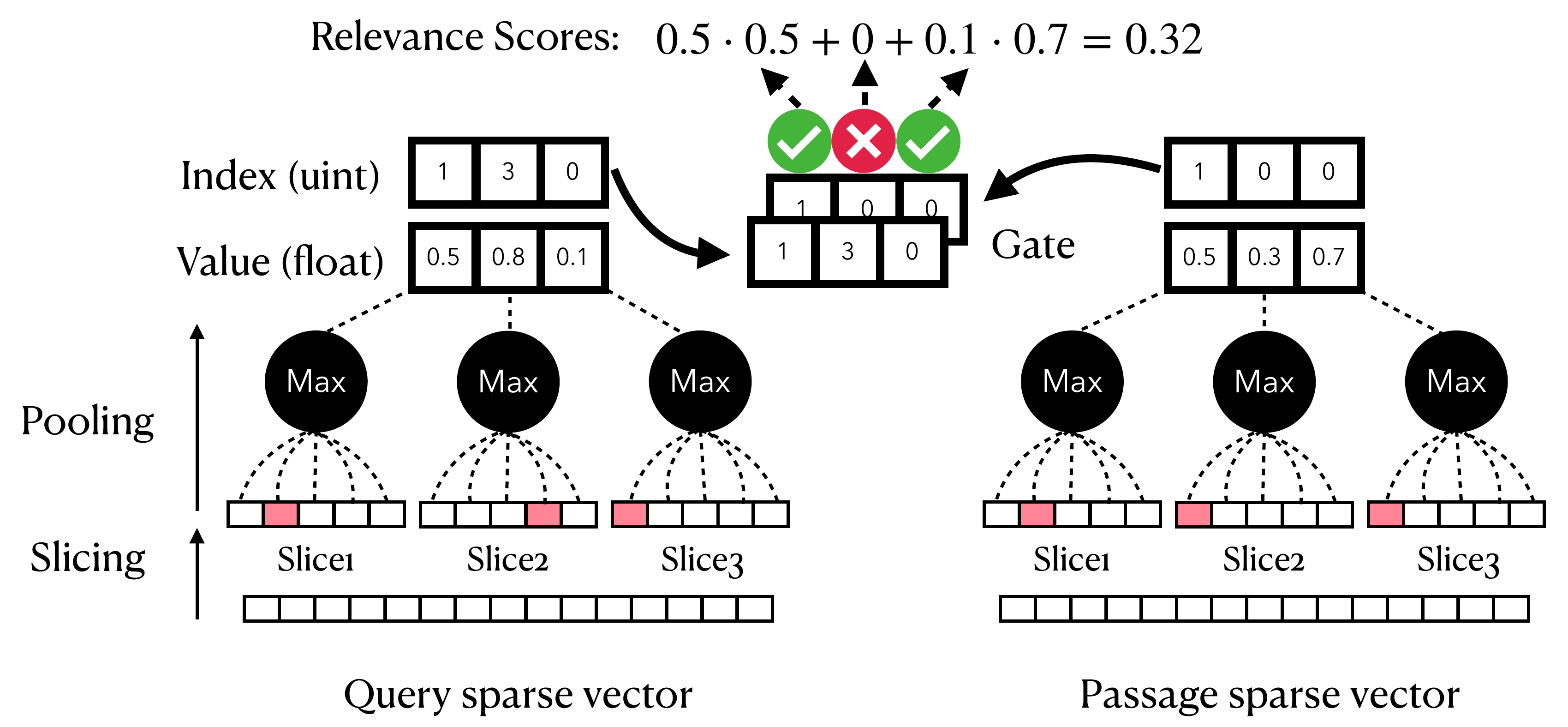}
    }
    \caption{With sparse representations, both queries and passages are represented as fixed-width vectors where the number of dimensions is equal to the vocabulary size.  Our approach first groups these high-dimensional vectors into $M$ slices, each with $N$ dimensions (e.g., $M=3$, $N=5$ here). For each slice, the maximum value is selected. These values from the original vector (Value) and their positions in each slice (Index) are recorded separately. When computing the gated inner product (GIP) between two vectors, only the dimensions with the same index are considered.
    }
    \label{fig:idea}
    \vspace{-0.2cm}
\end{figure}

Recently, another thread of work uses bi-encoders to learn sparse representations for text retrieval.
For example, \citet{deepct} demonstrate that replacing tf--idf with contextualized term weights via a regression model significantly improves retrieval effectiveness. \citet{deepimpact} further improve the approach by combining contextualized term weights with passage expansion models based on sequence-to-sequence transformers \citet{doctttttquery}, addressing the vocabulary mismatch problem with sparse retrieval.

Around the same time, \citet{splade, tilde} propose bi-encoder designs for jointly learning contextualized term weights and expansions.
These papers demonstrate that sparse retrievers generate representations that are more understandable for humans than dense retrievers.
Furthermore, there is evidence~\cite{mebert, clear, unicoil} that sparse representations compensate for the weaknesses of dense representations; these researchers fuse sparse and dense representations for text retrieval. \citet{spar} recently propose to train another dense retriever (SPAR) to imitate sparse retrievers, and directly fuse the dense representations from SPAR and DPR.

The advantages of sparse retrievers motivate us to explore the following research question: Can we densify sparse representations?
Our motivation comes from two reasons:\ (1) The densified representations are more likely to be interpretable and can serve as a good alternative to dense representations in scenarios where interpretability is important.
(2) The densified representations can be directly combined with other dense representations under the same physical retrieval framework, which makes these techniques easier to deploy. 

We present an approach to densifying sparse representations by representational slicing.
Unlike previous work~\cite{spar}, our approach involves no training. 
In addition, we propose a gated inner product (or GIP) operation to compute the relevance scores between the densified sparse representations (DSRs) of queries and passages. 
This process is depicted in Fig.~\ref{fig:idea}.
Although our analysis indicates that GIP has higher time complexity compared to inner products for dense vectors, with our proposed retrieve-and-rerank approach, DSR retrieval can be further sped up without any retrieval effectiveness drop.
Finally, our experiments show that combining DSRs with other dense representations yields performance tradeoffs that balance effectiveness, index size, and query latency.

\section{Background}

Following \citet{lin2020pretrained}, let us formulate the task of text (or \textit{ad hoc}) retrieval as follows:\ Given a query $q$, the goal is to retrieve a ranked list of documents $\{d_1, d_2, \cdots d_k\} \in C$ to maximize some ranking metric, such as nDCG, where $C$ is the collection of documents.  

Specifically, given a (query, document) pair, we aim to maximize the following:
\begin{align}
    P(\text{Relevance}|q, d) \triangleq \phi(\eta_q(q), \eta_d(d)) =  \phi(\mathbf{q}, \mathbf{d}) \nonumber
\end{align}
\noindent where $\eta_q(q)$ and $\eta_d(d) \in \mathbb{R}^h$ denote the functions mapping the query and document into $h$-dimensional vector representations, $\mathbf{q}$ and $\mathbf{d}$, respectively.
The function that quantifies the degree of relevance between the representations $\mathbf{q}$ and $\mathbf{d}$ is denoted $\phi(\cdot,\cdot)$, which can be a simple inner product or a much more complex operation~\cite{marco_BERT, poly-encoders}. 

\paragraph{Dense retrieval.} Transformer-based bi-encoders have been widely applied to the task of passage retrieval~\cite{sentence-bert, Chang2020Pre-training, dpr}.
The query and passage texts are first projected to low-dimensional dense vectors, and their inner product is computed as a measurement of relevance:
\begin{align}
    P^{ds}(\text{Relevance}|q, d) \triangleq  \langle \mathbf{q}^{\texttt{[CLS]}}, \mathbf{d}^{\texttt{[CLS]}} \rangle \nonumber, 
\end{align}
where $\mathbf{q}^{\texttt{[CLS]}}$ and $\mathbf{d}^{\texttt{[CLS]}}\in \mathbb{R}^{768}$ are the query and passage dense representations encoded by the [CLS] token from the final layer of BERT-base.

\paragraph{Sparse retrieval.} \citet{SNRM} was the first to demonstrate that neural networks can learn sparse representations for text retrieval.
Recently, transformer-based bi-encoders have also been applied to sparse representation learning.
Briefly, the solutions can be classified into two approaches.
On the one hand, some researchers~\cite{deepct,deepimpact,tilde,unicoil} use BERT token embeddings to learn a contextualized term weight for each input token.
On the other hand, some researchers~\cite{splade, sparterm} design models to directly learn sparse representations on the entire BERT wordpiece vocabulary.

Generally, all these sparse retrieval approaches can be considered as projecting queries and passages into $|V_{\text{BERT}}|$-dimensional vectors, where $|V_{\text{BERT}}|=30522$ is the vocabulary size of BERT wordpiece tokens:
\begin{align}
    P^{sp}(\text{Relevance}|q, d) \triangleq  \langle \mathbf{q}^{sp}, \mathbf{d}^{sp} \rangle \nonumber, 
\end{align}
where $\mathbf{q}^{sp}$ and  $\mathbf{d}^{sp}\in \mathbb{R}^{30522}$.
The value in each dimension is the token's term weight. 
The same as dense retrieval, the relevance score between a query and a passage is computed by the inner product of their vector representations.

\paragraph{Gap between dense and sparse retrieval.}
Although the inner product is the common operation for measuring relevance for both dense and sparse representations, there are still some major differences between them. 
(1) Unlike dense representations, sparse ones can be considered as bag of words and thus are more interpretable, since each dimension of the vectors represents a word in the BERT vocabulary.
(2) Text retrieval using sparse representations are executed through inverted indexes due to their high dimensionality, while dense retrieval is often executed on flat indexes using Faiss~\cite{faiss}. 
This gap between the execution of dense and sparse retrieval increases the complexity of dense--sparse hybrid retrieval systems.
Previous work~\cite{mebert, clear, unicoil} realizes hybrid retrieval by combining dense and sparse retrieval scores from different retrieval systems, which requires further post processing.
Also, it is not trivial to perform dense and sparse retrieval (i.e., on inverted and Faiss flat indexes) simultaneously. 
Thus, we are motivated to explore approaches to densifying sparse representations.

\begin{figure}[t]
    \centering
    \resizebox{\columnwidth}{!}{
        \includegraphics{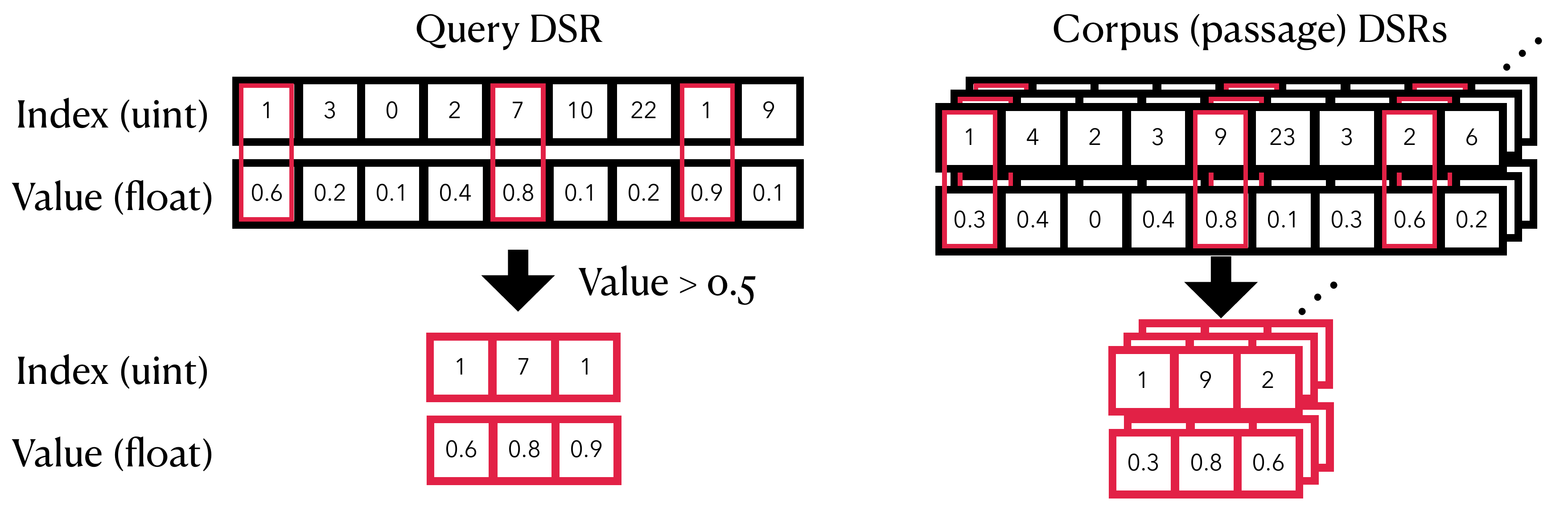}
    }
    \caption{DSR retrieval can be approximated by only computing the gated inner product from a few dimensions where the query value vector has large values (e.g, greater than 0.5) and then precisely reranking the top-$k$ passages using all the dimensions.}
    \label{fig:two-stage}
    \vspace{-0.2cm}
\end{figure}

\section{Methodology}
\label{sec:method}

In this section, we first introduce our approach to densifying sparse representations into low-dimensional dense vectors.
We also propose a simple scoring model for text retrieval with the densified sparse representations (DSRs).

\subsection{Densify Sparse Representations}
Sparse representations can be considered vectors with $|V|$ dimensions, i.e., $\mathbf{q}^{sp}=(q_0,\cdots q_{|V|-1})$; $\mathbf{d}^{sp}=(d_0, \cdots d_{|V|-1})$.
We first divide the vectors into $M$ slices of smaller vectors,  each of which has $N$ dimensions (i.e., $|V|=M \cdot N$).
In terms of a standard ``slice'' notation:
\begin{align}
    &S^{q}_m = \mathbf{q}^{sp}[mN:mN+N] \in \mathbb{R}^{N} \\ 
    &S^{d}_{m} = \mathbf{d}^{sp}[mN:mN+N] \in \mathbb{R}^{N}, 
\end{align}
\noindent where $m \in \{0,1,\cdots,M-1 \}$.
Note that the slicing can be performed in different ways (e.g., contiguous, stride,\footnote{The array is sliced with $N$ stride.} or random); for simplicity, we use contiguous in our presentation.
Thus, the inner product between $\mathbf{q}^{sp}$ and $\mathbf{d}^{sp}$ can be rewritten as the summation of all their slices' dot products:
\begin{align}
\label{eq:bucket_dot}
    \langle\mathbf{q}^{sp}, \mathbf{d}^{sp}\rangle =  \sum_{m=0}^{M-1} \langle S^{q}_m , S^{d}_m \rangle
\end{align}


\paragraph{Approximate sparse representations.} Intuitively, if a representation is sparse enough, we can assume that for each slice, there is only one non-zero entry.
Thus, we can approximate $S^{q}_m$ ($S^{d}_m$) by keeping only the entry with maximum value in each slice:    
\begin{align}
    S^{q}_m \approx \max{(S^{q}_m)} \cdot \hat{\mathbf{u}}(\text{e}^{q}_m) \\
     S^{d}_m \approx \max{(S^{d}_m)} \cdot \hat{\mathbf{u}}(\text{e}^{d}_m),
\end{align}
where $\hat{\mathbf{u}}(\text{e}^q_m)$ is a unit vector with the only non-zero entry at the entry $\text{e}^{q}_{m}=\mathrm{argmax}{(S^{q}_{m})}$.
Thus, the inner product of $\mathbf{q}^{sp}$ and $\mathbf{d}^{sp}$ sparse vectors in Eq.~(\ref{eq:bucket_dot}) can be approximated as follows:
\begin{align}
\label{eq:approximate_dot}
    &\langle\mathbf{q}^{sp}, \mathbf{d}^{sp}\rangle \approx \nonumber \\
    &\sum_{m=0}^{M-1}  \max{(S^{q}_m)} \cdot \max{(S^{d}_m)} \cdot \langle \hat{\mathbf{u}}(\text{e}^{q}_{m}) , \hat{\mathbf{u}}(\text{e}^{d}_{m}) \rangle \nonumber\\
    &= \sum_{m=0}^{M-1}  \max{(S^{q}_m)} \cdot \max{(S^{d}_m)} \mathds{1}_{\{ \text{e}^{q}_{m}= \text{e}^{d}_{m} \}}
\end{align}

\paragraph{Densified sparse representations.}
Observing Eq.~(\ref{eq:approximate_dot}), in order to compute the approximate inner product of sparse vectors, each query (document) can be alternatively represented as two $M$-dimension dense vectors:
\begin{align}
    \mathbf{q}^{ds} &= (\max{(S^{q}_{0})},\cdots,\max{(S^{q}_{M-1})}) \in \mathbb{R}^{M}  \nonumber \\
    \mathbf{d}^{ds} &= (\max{(S^{d}_{0})},\cdots,\max{(S^{d}_{M-1})}) \in \mathbb{R}^{M} \nonumber \\
     \mathbf{e}^{q} &= (\text{e}^{q}_{0},\cdots,\text{e}^{q}_{M-1}) \in \mathbb{N}^{M} \nonumber \\
    \mathbf{e}^{d} &= (\text{e}^{d}_{0},\cdots,\text{e}^{d}_{M-1}) \in \mathbb{N}^{M}, \nonumber 
\end{align}
where $\mathbf{q}^{ds}$ ($\mathbf{d}^{ds}$) is the dense vector storing the $M$ maximum values from the query (document) slices, and $\mathbf{e}^{q}$ ($\mathbf{e}^{d}$) is the integer dense vector storing the entries with the maximum value in the corresponding slices.
We call the representations $\mathbf{q}^{ds}$ and $\mathbf{e}^{q}$ ($\mathbf{d}^{ds}$ and $\mathbf{e}^{d}$) the \textit{densified sparse representations} (DSRs) for queries (documents).



\paragraph{Gated inner product.}
Using DSRs, Eq.~(\ref{eq:approximate_dot}) can be rewritten as:
\begin{align}
\label{eq:approximate_dot1}
    &\langle\mathbf{q}^{sp}, \mathbf{d}^{sp}\rangle \approx \nonumber \\ &\sum_{m=0}^{M-1}  \mathbf{q}^{ds}[m] \cdot \mathbf{d}^{ds}[m] \mathds{1}_{\{ \mathbf{e}^{q}[m]= \mathbf{e}^{d}[m] \}},
\end{align}
where $\mathbf{q}^{ds}[m]$ ($\mathbf{d}^{ds}[m]$) and $\mathbf{e}^{q}[m]$ ($\mathbf{e}^{d}[m]$) is the $m$-th entry of query (document) DSRs.
We call the operation in Eq.~(\ref{eq:approximate_dot1}) the \textit{gated inner product} (GIP), which can be considered an approximation of the inner product of the original sparse vectors. 

However, note that the gated inner product requires $4 \cdot M$ operations, which is more than the standard inner product of $M$-dimensional dense vectors, which only requires $2 \cdot M$ operations.
When conducting brute-force search over corpus $C$, the differences are even larger:\ $4 \cdot M |C|$ > $2 \cdot M|C|$.

\paragraph{Retrieval and reranking.}
In order to address the issue of high time complexity, we propose a retrieve-and-rerank approach.
In the first stage, we use approximate brute-force search by computing only partial dimensions of the gated inner product between the query and document DSRs:
\begin{align}
\label{eq:approximate_retrieval}
    &\sum_{m \in \mathcal{M}}  \mathbf{q}^{ds}[m] \cdot \mathbf{d}^{ds}[m] \mathds{1}_{\{ \text{e}^{q}[m]= \text{e}^{d}[m] \}},
\end{align}
where $\mathcal{M} = \{ m |\mathbf{q}^{ds}[m] > \theta \}$ is the set of indices for the GIP computation, and $\theta$ is a hyperparameter.
This first-stage retrieval only relies on the dimensions where  $\mathbf{q}^{ds}[m]$ has a large value.
The intuition behind this design is that user queries usually contain only a few terms.
Thus, only a few dimensions in the sparse query vector have impact on retrieval effectiveness, which might be also true for the DSRs. 
In the second (reranking) stage, the retrieved documents are sorted by the approximate gated inner product; then, the top-$k$ documents are reranked by gated inner product with all dimensions, as in Eq.~(\ref{eq:approximate_dot1}).

\begin{table*}[t]
	\caption{Results of sparse representation densification experiments.}
	\label{tb:dim_reduction}
	\vspace{-0.2cm}
	\centering
	\small
	\begin{threeparttable}
    \begin{tabular}{lccccccc}
	\toprule
	&  &\multicolumn{2}{c}{Dev}& \multicolumn{2}{c}{DL 2019}&\multicolumn{2}{c}{DL 2020}\\ 
\cmidrule(lr){3-4} \cmidrule(lr){5-6}\cmidrule(lr){7-8}
	& M (Dims) & MRR@10&  R@1K& nDCG@10& R@1K& nDCG@10& R@1K \\ 
\midrule
      & 30K  & 0.312 & 0.925 & 0.621  & 0.775 & 0.633 & 0.794  \\ 
  DSR-uniCOIL   & 768   & 0.309 & 0.923 & 0.615  & 0.769 & 0.632 & 0.792 \\ 
       (no expansion) &  256   & 0.305 & 0.919 & 0.605 & 0.751 & 0.622 & 0.788 \\ 
       & 128    & 0.300 & 0.913 & 0.605 & 0.744 & 0.614 & 0.782  \\ 
       \midrule
     \multirow{4}{*}{DSR-SPLADE}   & 30K\tnote{a}  & 0.340 & 0.965 & 0.684 & 0.851 & - & - \\ 
       & 768  & 0.343 & 0.957 & 0.681 & 0.831 & 0.655 & 0.824 \\ 
       & 256  & 0.338  & 0.953 & 0.667 & 0.820 & 0.651 & 0.822 \\ 
       & 128 & 0.332  & 0.946 & 0.657 & 0.810 & 0.657 & 0.812 \\ 
       
	\bottomrule
	\end{tabular}
	\begin{tablenotes}
	    \item[a] These numbers are copied from SPLADE-max in~\citet{splade-v2} for reference only and are {\it not} directly comparable to our results.
    \end{tablenotes}
    \end{threeparttable}
\end{table*}

\begin{figure*}[h]
\centering
\begin{subfigure}[t]{.48\textwidth}
    \includegraphics[width=\columnwidth]{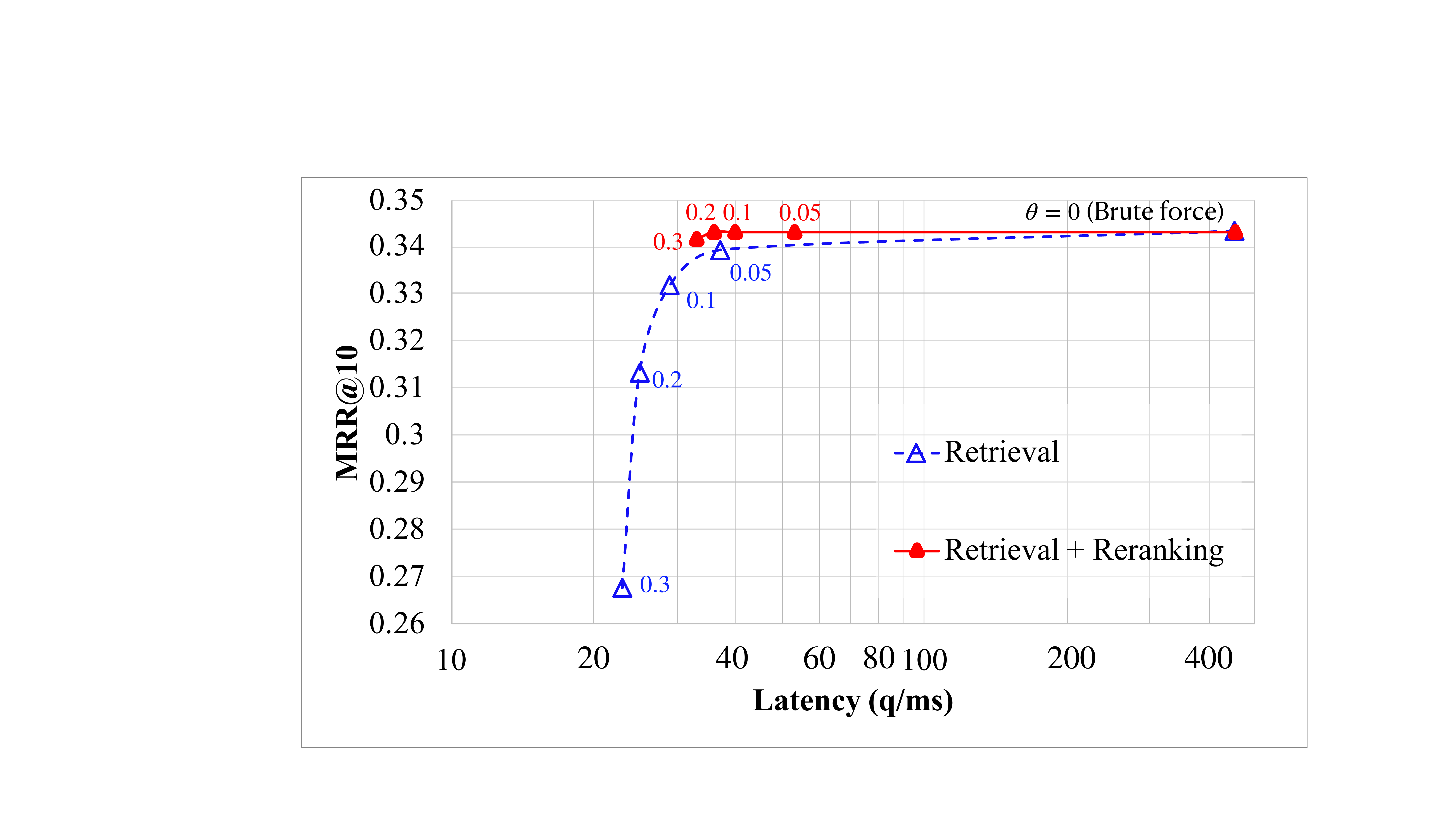}
    \caption{MRR@10}
\end{subfigure}
\begin{subfigure}[t]{.48\textwidth}
    \includegraphics[width=\columnwidth]{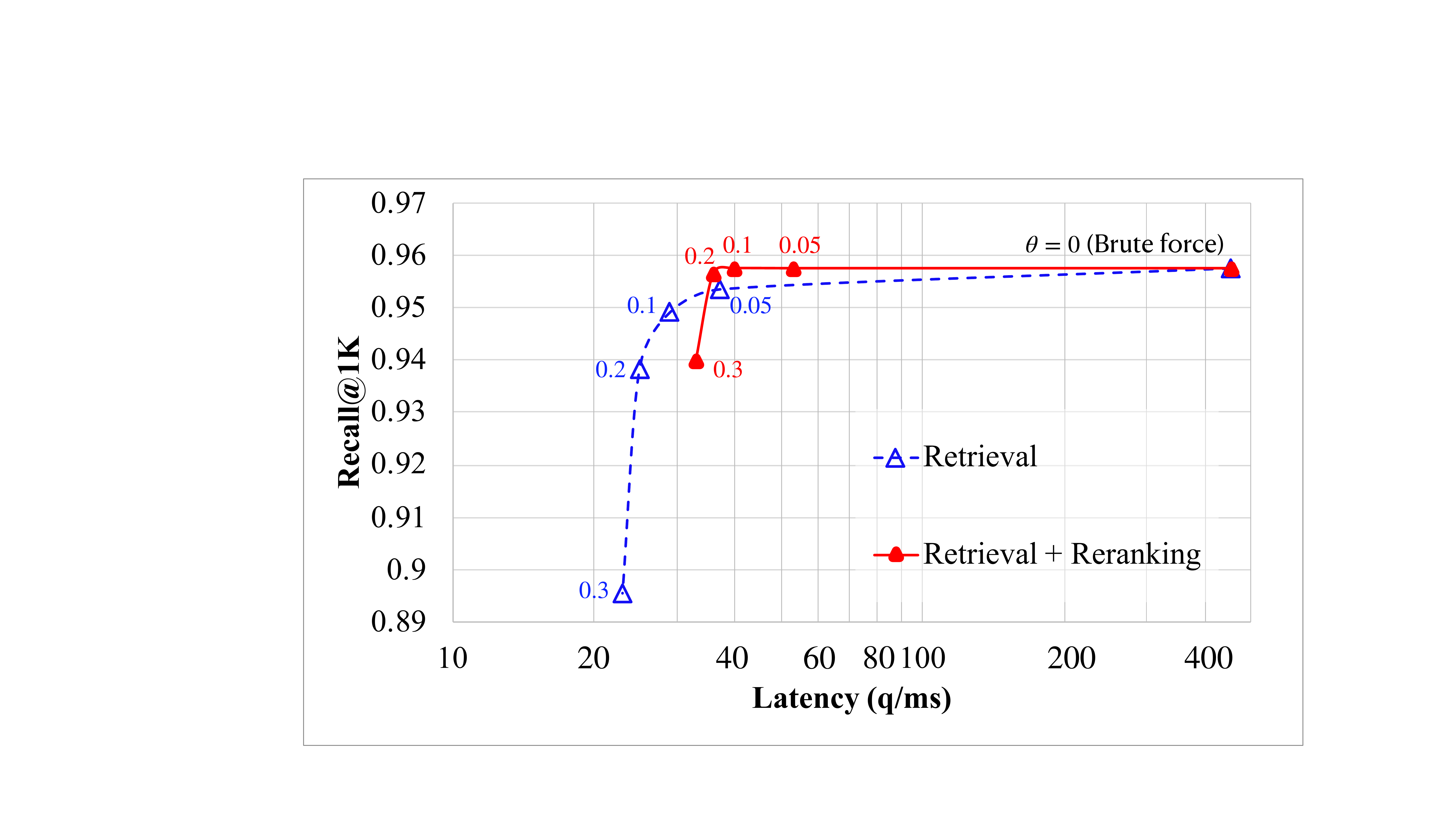}
    \caption{Recall@1K}
\end{subfigure}
\caption{Experiments with our retrieve-and-rerank approach using DSR-SPLADE (768 dimensions) under different $\theta$ settings. $\theta=0$ means brute force search using all dimensions.}
\label{fig:fast_retrieval}
\end{figure*}

\begin{table*}[h]
	\caption{Effectiveness/efficiency comparisons.}
	\label{tb:main_result}
	\vspace{-0.2cm}
	\centering
	 \resizebox{0.95\textwidth}{!}{  
	 \begin{threeparttable}
    \begin{tabular}{lcccccrr}
	\toprule
&\multicolumn{2}{c}{Dim}&Dev & DL 2019 & DL 2020 & storage  & latency \\
	\cmidrule(lr){2-3} \cmidrule(lr){4-4} \cmidrule(lr){5-5} \cmidrule(lr){6-6}
&	Sparse & Dense &  MRR@10&  nDCG@10 &  nDCG@10 & (GB) & (ms/q) \\
\midrule
      ColBERT~\cite{colbert} & 0 & 128/tok & 0.360 & - & - & 154 & 458\tnote{a}\\
      COIL~\cite{coil} & 32/tok & 768 & 0.355 & 0.704 & - & 50 & 41\tnote{a}\\
      Dense-[CLS] (our dense baseline) & 0 & 768 &  0.310 & 0.626 & 0.625 & 13 & 115 \\
        \midrule
        DSR-uniCOIL  & 768 & 0 & 0.309 & 0.615  & 0.632 & \multirow{2}{*}{20} & \multirow{2}{*}{40} \\
        DSR-SPLADE  & 768 & 0 & 0.343 & 0.681 & 0.655  & &  \\
    \midrule
        DSR-uniCOIL + Dense-[CLS] & 768 & 768  & 0.350  & 0.678  & 0.704  & \multirow{2}{*}{32} & \multirow{2}{*}{2 $\times$ 24\tnote{b}} \\
        DSR-SPLADE  + Dense-[CLS]& 768 & 768 & 0.352 & 0.697  & 0.677  &   &  \\
    \midrule
        DSR-uniCOIL + Dense-[CLS] & 256 & 256  & 0.346  & 0.680  & 0.688  & \multirow{2}{*}{11} & \multirow{2}{*}{34} \\
        DSR-SPLADE  + Dense-[CLS]& 256 & 256 & 0.348 & 0.711  & 0.678 &   &  \\
    \midrule
        DSR-uniCOIL + Dense-[CLS]& 128 & 128   & 0.337  & 0.679  & 0.671 &  \multirow{2}{*}{5} & \multirow{2}{*}{32} \\
        DSR-SPLADE  + Dense-[CLS]& 128 & 128   & 0.344 & 0.709 & 0.673 &   &  \\
	\bottomrule
	\end{tabular}
	\begin{tablenotes}
	    \item[a] These numbers are copied from the original papers, with different execution environments.
	    \item[b] This condition cannot run on our single GPU; thus, we divide the index into 2 shards, each with a latency of 24 ms.
    \end{tablenotes}
    \end{threeparttable}
	}
\end{table*}

\subsection{Fusion with Dense Representations}

One of the advantages of densifying sparse representations is to make the fusion of sparse and dense retrieval easier, since it actually becomes the fusion of two dense representations.
Their fusion scores can be computed as follows: 
\begin{align}
\label{eq:combine}
 \sum_{m=0}^{M-1}  &\lambda \cdot \mathbf{q}^{\texttt{[CLS]}}[m] \cdot \mathbf{d}^{\texttt{[CLS]}}[m] \nonumber\\
 &+ \mathbf{q}^{ds}[m] \cdot \mathbf{d}^{ds}[m] \mathds{1}_{\{ \mathbf{e}^{q}[m]= \mathbf{e}^{d}[m] \}},
\end{align}
\noindent where $\lambda$ is a hyperparameter. Note that the sparse and dense representations can be from the same model using joint training or from separately trained models. Eq.~(\ref{eq:combine}) can be implemented by existing packages that support common array operations.
We use PyTorch in our experiments.

\section{Experimental Setup}

\paragraph{Dataset descriptions.}
(a) MS MARCO Dev:\ 6980 queries comprise the development set for the MS MARCO passage ranking test collection, with on average one relevant passage per query.
We report MRR@10 and R@1000 as the top-$k$ retrieval measures.
(b) TREC DL~\cite{trec19dl, trec20dl}:\ the organizers of the 2019 (2020) Deep Learning Track at the Text REtrieval Conferences (TRECs) released 43 (53) queries with multiple graded relevance labels, where (query, passage) pairs were annotated by NIST assessors.
We report NDCG@10 and R@1000 for these evaluation sets.

\paragraph{Sparse retrieval models.}
In our experiments, we explore two previous sparse retrieval models based on BERT-base: 

\begin{enumerate}[leftmargin=*]
\item uniCOIL~\cite{unicoil} uses token embeddings from the final layer in BERT to generate one-dimension representations (term weights) for each input query and passage token.
In our experiments, we use uniCOIL without passage expansion for simplicity.

\item SPLADE~\cite{splade} uses the MLM technique to project each token embeddings from query (or passage) to a $|V_{\text{BERT}}|$-dimensional sparse representation, where $|V_{\text{BERT}}|=30522$ is the vocabulary size of BERT wordpiece.
Following \citet{splade-v2}, we conduct max pooling over all sparse representations generated by query (or passage) tokens. 
Note that unlike \citet{splade-v2}, we do not control the sparsity of generated representations while training.
\end{enumerate}

\noindent We denote the densified sparse representations from uniCOIL (SPLADE) as DSR-uniCOIL (DSR-SPLADE).
We train all models on the MS MARCO ``small'' triples training set for 100k steps with a batch size of 96 and learning rate 7e-6.
In our experiments, we test retrieval latency using single NVIDIA Titan RTX with batch size 1.

We densify the 30522-dimensional representations into 768-dimensional vectors by (1) discarding the first 570 unused tokens in the BERT vocabulary; (2) divide the remaining 29952 tokens into 768 slices.
Each of the slices contains 39 tokens according to token ids; i.e., $M=768$ and $N=39$.
In addition, we also conduct experiments for densifying into 256 and 128 dimensions, where there are 117 and 234 tokens in each slice, respectively. The ``value'' and ``index'' dense vectors (see Fig.~\ref{fig:idea}) are stored as float16 and uint8, respectively.

\section{Results}

\subsection{Densifying Sparse Representations}

Table~\ref{tb:dim_reduction} shows our results for densifying sparse representations.
The first row (with 30K dimension) in each block reports the retrieval effectiveness of the original sparse representations, which can be considered an upper bound.
Note that uniCOIL's representation is highly sparse, and thus retrieval can be performed using an inverted index.
However, since our SPLADE model is trained without any sparsity constraints, the output query and passage representations are actually quite dense and challenging for end-to-end retrieval; see~\citet{Mackenzie_etal_arXiv2021} for more discussion.
For reference, we report the retrieval effectiveness of the SPLADE-max model from the original paper~\cite{splade-v2}, which {\it does} include sparsification that renders the model amenable to inverted indexes.

As we can see from Table~\ref{tb:dim_reduction}, our method is able to densify the original 30K-dimensional sparse vectors into 768-dimensional vectors with less than $1\%$ retrieval effectiveness drop.
In addition, only around $2\%$ and $4\%$ retrieval effectiveness degradation can be seen for $M=256$ and $128$, respectively.
These results demonstrate the advantage of our approach, which does not require any training.

\subsection{Retrieve and Rerank}
\label{subsec:retrieval-exp}

In Section~\ref{sec:method}, we discussed the disadvantage of the gated inner product operation, that it is more expensive than the standard inner product. 
In these experiments, we test SPLADE with 768 dimensions on the MS MARCO development queries to demonstrate how our proposed retrieve-and-rerank approach remedies the issue. 

In Fig.~\ref{fig:fast_retrieval}, the red lines show the effectiveness and efficiency of the retrieve-and-rerank approach under different settings of $\theta$ (which controls the value threshold in GIP), while the blue lines depict first-stage retrieval only.
Note that the $x$ axis is in log scale.
The point $\theta=0$ represents the retrieval effectiveness upper bound of the model since the gated inner product is computed using all dimensions.
As $\theta$ increases, the first-stage (approximate) retrieval efficiency improves dramatically but effectiveness drops.
However, we see no effectiveness drops with $\theta \leq 0.1$ after reranking the top-10K candidates from first-stage retrieval, which only consumes an additional latency of 10 ms.
Thus, in the following experiments, we set $\theta$ to $0.1$.

\subsection{Comparisons to Other Dense Models}

Table~\ref{tb:main_result} compares the performance of DSR and other dense retrievers in terms of retrieval effectiveness, latency, and index size.
For a fair comparison, here we exclude other dense models trained with more sophisticated techniques, such as knowledge distillation and hard negative mining. 
We implement a bi-encoder model (denoted ``Dense-[CLS]'') trained under the same condition as our sparse models; this model takes the [CLS] token as the output representation.
Note that the latency of ``Dense-[CLS]'' is measured using a Faiss flat index.

In the conditions denoted DSR-\{uniCOIL, SPLADE\} + Dense-[CLS], we train a bi-encoder model for sparse and dense representations {\it jointly}, where the loss takes into account inner products from both the [CLS] token and the sparse vectors.
After training, we then densify the sparse model as described above.
During inference, we apply our retrieve-and-rerank approach, where 10K candidates are first retrieved by only DSRs and then reranked by the fusion of DSRs and dense representations, as in Eq.~(\ref{eq:combine}).
Here we set $\lambda$ to 1.

Among the dense retrievers considered here, ColBERT is the most effective, and COIL, which can be thought of as a variant of ColBERT, reduces the time complexity by adding a lexical match prior. 
However, both multi-vector designs still require much more storage for the index than single-vector dense retrievers. 
Our fusion models combining 768-dimensional DSRs with dense [CLS] representations show retrieval effectiveness close to COIL but with a much smaller index size.
When reducing the dimensions of the DSR models, we can see a tradeoff between efficiency and effectiveness. 
However, it is worth noting that our fusion DSR models with 128-dimensional vectors shows better efficiency and effectiveness than the dense-only and DSR-model.     
This shows that our densified sparse representations exhibit complementary relevance signals than ``pure'' dense retrieval models.

\begin{table}[t]
	\caption{Ablation on slicing method.}
	\label{tb:ablation}
	\vspace{-0.2cm}
	\centering
	\small
    \begin{tabular}{lcc}
	\toprule
	&\multicolumn{2}{c}{MS MARCO dev} \\
	\cmidrule(lr){2-3}
	DSR-SPLADE &  MRR@10&  R@1K \\
\midrule
         Stride & 0.343 & 0.957 \\
         Contiguous & 0.332 & 0.951 \\
         Random & 0.343 & 0.957 \\
	\bottomrule
	\end{tabular}
\end{table}

\subsection{Alternative Slicing Strategies}

Finally, we examine different slicing strategies for densifying sparse representations using DSR-SPLADE with 768 dimensions in these experiments. 
Table~\ref{tb:ablation} reports the results.
Our default setting for the previous experiments is stride and we observe that stride has the same effectiveness as randomized slicing. 
Surprisingly, the contiguous slicing strategy shows degradation in ranking effectiveness. 
This indicates that max pooling over contiguous slices could lose information compared to the other strategies.
It is worth exploring other slicing strategies, which we leave for future work.

\section{Conclusions}

In this paper, we study how to densify sparse representations for passage retrieval.
Our approach requires no training and can be considered an approximation of sparse representations.
The relevance score of the densified sparse representations (DSRs) can be computed through our proposed gated inner product (GIP), a variant of the standard inner product.
Although our analysis indicates that GIP has higher time complexity than inner product, our experiments show that in practice, DSR retrieval can be conducted through a retrieve-and-rerank approach, which is 10 times faster without sacrificing any effectiveness.
Finally, we empirically show that a bi-encoder model can be a more robust dense retriever by combining DSRs along with other ``standard'' dense representations.

\section*{Acknowledgements}

This research was supported in part by the Canada First Research Excellence Fund and the Natural Sciences and Engineering Research Council (NSERC) of Canada.
Additionally, we would like to thank the support of Cloud TPUs from Google's TPU Research Cloud (TRC).

\bibliographystyle{acl_natbib}
\bibliography{paper.bib}


\end{document}